\documentclass[letterpaper,conference]{IEEEtran}

\usepackage[compress]{cite}
\usepackage{subfigure} 
\usepackage{amsmath,amssymb,amsfonts,amsthm}
\usepackage{textcomp}
\usepackage{xcolor}
\usepackage{soul} 
\usepackage{booktabs} 
\usepackage{makecell} 
\usepackage{pgf,pgfkeys,tikz,circuitikz,float,bm,caption}
\usepackage[ruled,vlined]{algorithm2e}
\usepackage{graphicx,multirow}%
\usepackage[super]{nth}
\usepackage{steinmetz}
\usepackage{siunitx,upgreek}
\usepackage{orcidlink}
\def\BibTeX{{\rm B\kern-.05em{\sc i\kern-.025em b}\kern-.08em
		T\kern-.1667em\lower.7ex\hbox{E}\kern-.125emX}}

\DeclareSIUnit{\rad}{rad}

\DeclareMathOperator{\Exp}{\mathbb{E}}

\graphicspath{{figures/}}

\begin{document}
	
\title{RFI and Jamming Detection in Antenna Arrays with an LSTM Autoencoder}




\author{Christos Ntemkas and Antonios Argyriou\\
Department of Electrical and Computer Engineering, University of Thessaly, 38334 Volos, Greece.}


\maketitle

\IEEEpubidadjcol

\begin{abstract}
Radio frequency interference (RFI) and malicious jammers are a significant problem in our wireless world. Detecting RFI or jamming is typically performed with model-based statistical detection or AI-empowered algorithms that use an input baseband data or time-frequency representations like spectrograms. In this work we depart from the previous approaches and we leverage data in antenna array systems. We use Fourier imaging to localize spatially the sources and then deploy a deep LSTM autoencoder that detects RFI and jamming as anomalies. Our results for different power levels of the RFI/jamming sources, and the signal of interest, reveal that our detector offers high  performance without needing any pre-existing knowledge regarding the RFI or jamming signal.
\end{abstract}




\section{Introduction}

Radio frequency interference (RFI) and malicious RF jamming are significant problems for a plethora of applications. RF sensing systems must detect and classify these types of signals so that appropriate counter-measures can be implemented. Detection can be realized with model-based algorithms that incorporate knowledge regarding the undesired signal that must be detected, e.g., cyclo-stationarity for RFI that consists of communication signals~\cite{gardner1986}. Since RFI is typically assumed to originate from wireless communication, modulation classification with Artificial Intelligence (AI) techniques~\cite{oshea2018,varkatzas2023}, can be re-purposed for RFI detection. By using features like spectrograms~\cite{varkatzas2023,yu2020}, higher order moments~\cite{oshea2018}, etc., AI can distinguish the unique behavior of RFI originating from wireless modulated signals and detect it. However, not all RFI consists of communication signals, or even signals of known behavior for which we can collect data and train AI models. Classifying these non-canonical signals where the injectors of RFI were malicious jammers was first considered in~\cite{karagiannis2018}, and later in~\cite{ghanney2020,yu2020,oyedare2023,singh2024}. Some of these systems may even exploit hardware impairments at the jammers for detection~\cite{argyriou2023a,arif2024}. In this work we show that for RFI or jammers to be detected, and potentially classified, we need to introduce data from additional sources, namely multiple antennas.

\begin{figure}[!htb]
	\centering
	\includegraphics[width=0.9\linewidth]{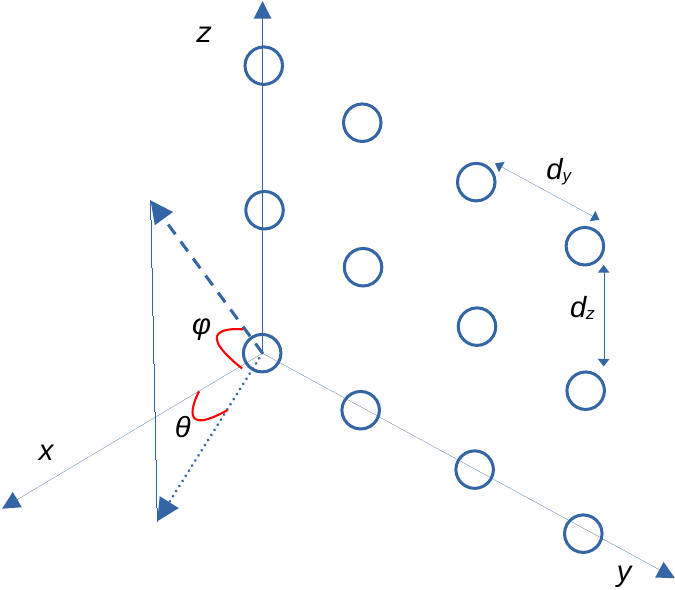}	
	\caption{URA geometry with the circles indicating the positions of the antennas in the $yz$ plane. The incoming signal arrives from direction $\theta,\phi$.}
	\label{fig:ura-geometry}
\end{figure}
\begin{figure}[t]
	\centering
	\includegraphics[width=0.9\linewidth]{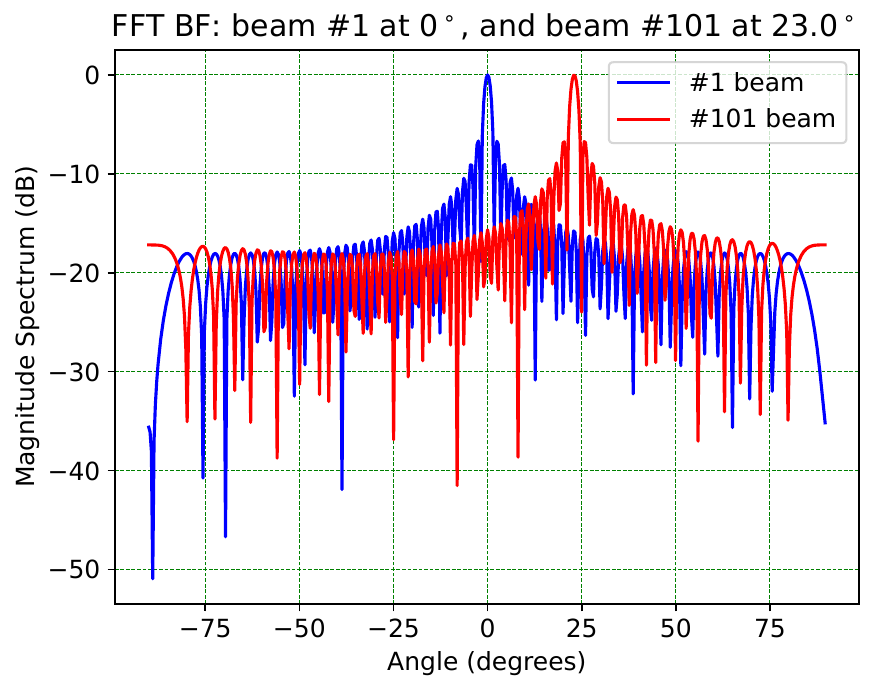}	
	\caption{Two beams obtained with DFT beamforming. We set $\phi=0$ and vary $\theta$.}
	\label{fig:1d-dft-beams}
\end{figure}

Our approach is pragmatic today due to the move in higher communication frequencies in modern communication systems (5G/6G). These systems use multiple antennas that are deployed to form array structures. Antenna arrays are used in radar, radio astronomy, and wireless communication, among other applications. Fig.~\ref{fig:ura-geometry} illustrates a 2D uniform rectangular array (URA). The purpose of an antenna array is to sample spatially an incoming RF wavefront. It can also be steered electronically to a desired spatial direction~\cite{vantrees2002}, something that can be accomplished with beamforming (BF) algorithms. The objective of BF is to maximize the output power of the array for signals incoming from a desired angle-of-arrival (AoA). This behavior is captured visually in the so-called \textit{array response} plot that can be 2D or 1D. A 1D cut of such a 2D array response can be seen in Fig.~\ref{fig:1d-dft-beams}. The first beam in this figure indicates that the signal originating from the desired \textit{look angle} of 0$^o$ will have the maximum power while signals from other AoAs will be attenuated.

In Fig.~\ref{fig:1d-dft-beams} signals from all other directions are considered as RFI or malicious jamming. Hence, we desire for them to be suppressed. Some beamforming techniques like the minimum variance distortionless response (MVDR), or Capon beamformer~\cite{capon1969}, have RFI suppression capabilities since they create nulls in the spatial direction of the RFI. But before RFI/jamming suppression schemes are deployed we must detect these events. 

In this work, we propose an RFI and jamming detection scheme that uses spatially-sampled data from an array that operates in passive receiving mode. We use these data 
to calculate the correlation matrix of the time-domain samples across the antennas and then calculate its discrete Fourier transform (DFT). As we will show next, when we have a 2D URA, the 2D-DFT of this correlation matrix provides an estimate of the signal power as a function of azimuth $\theta$ and elevation $\phi$. The result is an image of the sources present in the received signal versus their spatial location~\cite{richards2022}.
%
%
These images are used for training an AI system, namely a deep neural network (DNN), which in our case is selected to be a Long Short-Term Memory (LSTM) autoencoder. Besides the images, an additional feature we provide as input to the DNN is \textit{the desired look angle}. In the presence of RFI or jamming, the 2D-DFT images will contain peaks in spatial directions beyond the look angle. These are anomalies for which the autoencoder has not been trained and they can be detected.

\section{Data Model \& Assumptions}

\textbf{URA Model:} We consider a redundant array of antennas, and in particular a URA with a number of $N$$\times$$M$ elements (Fig.~\ref{fig:ura-geometry}). 
The unit vector oriented towards the directions of azimuth  $\theta$, elevation $\phi$ in Cartesian coordinates is $\mathbf{r_1}(\theta,\phi)=\cos(\phi)\cos(\theta)\mathbf{x_1}+\cos(\phi)\sin(\theta)\mathbf{y_1}+\sin(\phi)\mathbf{z_1}$, where $[\mathbf{x_1}~\mathbf{y_1}~\mathbf{z_1}]^T$ is the Cartesian unit vector.
%
Let us assume that the antennas are deployed along the $y,z$ dimensions, and are spaced apart by $d_y$ and $d_z$ meters, respectively. So its $x$ coordinate is zero. The location of the $n,m$-th element of the array is given by the Cartesian coordinates $\mathbf{r}=[nd_y~md_z]$, i.e., the Cartesian position vector is $\mathbf{r}_{n,m}=nd_y\mathbf{y_1}+md_z\mathbf{z_1}$.
For an incoming wave from direction $\phi,\theta$ that impinges on the array, the phase difference between the 0,0-th and the $m,n$-th element depends on the \textit{equivalent} distance $\mathbf{r}_{n,m}$$\cdot$$\mathbf{r_1}(\theta,\phi)$. For the $n,m$-th element we have:
\begin{align}
\mathbf{r}_{n,m}\cdot\mathbf{r_1}(\theta,\phi)=nd_y \cos(\phi)\sin(\theta)+md_z \sin(\phi)
\label{eqn:distance}
\end{align}
%
%
To calculate the impact of this geometry on the array data model, we have to discuss briefly some well-known but necessary aspects of the sensing instrument. Let the $MN\times 1$ vector $\mathbf{y}$ denote a single time domain sample of the baseband radio signal received at all the array elements after down-conversion to baseband and analog-to-digital conversion (ADC)~\cite{vanderveen2019}. When this signal is wideband with the help of a filterbank it is split into multiple narrowband frequency channels and the respective filtered baseband signals in each of these are then denoted as $\mathbf{y}(f)$. This is necessary so that the narrowband assumption holds and the effect of the geometrical delay across two elements of the array is captured only by a phase shift~\cite{vanderveen2019}. Then the phase shift between two elements is simply the angular wavenumber $2\pi/\lambda$ multiplied by their equivalent distance given in~\eqref{eqn:distance}.
Hence, the phase shift of the $n,m$-th element relative to the 0,0 is:
\begin{align}
	a_{n,m}(f,\phi,\theta)&=\exp \big ( j\frac{2\pi}{\lambda}(\mathbf{r}_{n,m}\cdot\mathbf{r_1}) \big  )\nonumber\\
	&=\exp ( j\frac{2\pi}{\lambda}(nd_y \cos(\phi)\sin(\theta)+md_z \sin(\phi)) )\nonumber\\
	& 0\leq m \leq M-1,~0\leq n \leq N-1
\end{align}
In the above expression, the spatial frequencies are $\cos(\phi)\sin(\theta)/\lambda$ and $\sin(\phi)/\lambda$ respectively.
With the narrowband assumption in mind, we can define the steering vector (array manifold) that captures the phase difference between the received signal at each element of the 2D array. It is the $MN\times 1$ vector $\mathbf{a}$ which is frequency dependent:
\begin{align}
	&\mathbf{a}^T =\Big [ 1~~\exp \big ( j\frac{2\pi}{\lambda}(1d_y \cos(\phi)\sin(\theta)+1d_z \sin(\phi)) \big )~...\nonumber\\
	&\exp \big (j\frac{2\pi}{\lambda}((N-1)d_y \cos(\phi)\sin(\theta)+(M-1)d_z \sin(\phi)) \big ) \Big ]
	\label{eqn:ura-steering-vector}
\end{align}

\textbf{Data Model for the URA:} The random signal of interest (SOI) when sampled is $x[l]$ and~\textit{is spatially narrow at the angles $\phi,\theta$}. $\sigma_x^2(\phi,\theta)=\Exp [|x[l]|^2]$ is the power of the SOI in the direction $\phi,\theta$. $z[l]$ is the sample of the overall RFI/jamming signal that also incorporates the steering vector (we do not model it like the SOI since it may come from multiple AoAs and is of no interest to process it).
The received signal for the $l$-th time instant and the $n,m$-th element is:
\begin{align}
	y_{n}[l]=a_{n,m}(f,\phi,\theta)x[l]+z[l]+w[l].
	\label{eqn:ura-signal-one-sensor}
\end{align}
In the above $w[l]$ denotes the Gaussian noise sample. The data model for the complete URA is again an $MN\times 1$ vector:
\begin{align}
\mathbf{y}[l]=	\mathbf{a}x[l]+\mathbf{z}[l]+\mathbf{w}[l].
	\label{eqn:vector-signal-model}
\end{align}


%
\textbf{Correlation:}
In theory the correlation matrix is:
\begin{align}
	\mathbf{R}_\mathbf{y}=\Exp [\mathbf{y}\mathbf{y}^H]=	\mathbf{a}\mathbf{a}^H\sigma_x^2(\theta,\phi)+\mathbf{R}_\mathbf{z}^2+\mathbf{R}_\mathbf{w}^2.
	\label{eqn:correlation}
\end{align} 
The dimensions of this matrix are $NM\times NM$ since it contains the correlation between a single element and all the remaining ones (including itself). It is important to understand precisely the contents of this matrix for designing later our detector.
Since we have a redundant array, this means that multiple antenna elements receive a signal coherently (the same phase) and so they can add them (coherent combining). This allows us to reach a simplified expression for each element of this matrix. Focusing on two antenna elements with coordinates $n_1,m_1$ and $n_2,m_2$, we set $n_1-n_2$=$k$, $m_1-m_2$=$\ell$ to capture the relative difference in the indexes of two elements. This difference is what matters for the calculation of correlation. Correlating the signal in \eqref{eqn:ura-signal-one-sensor} across these two antennas:
\begin{align*}
	&\Exp[y_{n_1,m_1}[l]y_{n_2,m_2}^*[l]]	=\nonumber\\
	&=\sigma^2_x(\theta,\phi)e^{j\frac{2\pi}{\lambda}((m_1-m_2)d_y \cos(\phi)\sin(\theta)+(n_1-n_2)d_z \sin(\phi)) }
	+\sigma^2_w
\end{align*}
If we add the the previous result across redundant antenna pairs (that satisfy $n_1-n_2$=$k$, and $m_1-m_2$=$\ell$):
\begin{align}	
r_{\ell,k}	&=\sum \limits_{\substack{\forall n_1-n_2=k \\ \forall m_1-m_2=\ell}} e^{j\frac{2\pi}{\lambda}(\ell d_y \cos(\phi)\sin(\theta)+kd_z \sin(\phi)) }\sigma^2_x(\theta,\phi)\nonumber\\
	&+\sigma^2_w, \ell \in [0,M],~k \in [0,N].\label{eqn:ura-correlation-interferometer}
\end{align} 
The last expression provides the theoretical values for each element of matrix \eqref{eqn:correlation}. In practice we can only estimate $\mathbf{R}_\mathbf{y}$: From the baseband data in \eqref{eqn:vector-signal-model} that we collect for each snapshot, we estimate the sample correlation matrix $\hat{\mathbf{R}}_\mathbf{y}$ as:
\begin{align}
	\hat{\mathbf{R}}_\mathbf{y}=\frac{1}{S}\sum_{s=1}^{S} \mathbf{y}_s\mathbf{y}_s^\textrm{H},
	\label{eqn:R_est}
\end{align} 
with $S$ being the number of collected snapshots of $\mathbf{y}$. So we essentially estimate each element of this matrix as $\hat{r}_{\ell,k}$. 

Note that our first goal is to estimate $\sigma^2_x(\theta,\phi)$ for every angle $\theta,\phi$ (which is what constitutes our images), by using $\hat{r}_{\ell,k}$ and~\eqref{eqn:ura-correlation-interferometer}. An efficient way to do this is with the DFT.

\section{Data Pre-Processing with 2D-DFT} 
Imaging with data from array samples uses the correlated data. The 2D-DFT of the correlations in~\eqref{eqn:ura-correlation-interferometer} provides the signal power over the different spatial directions~\cite{thompson2017,burke2019}. 
%
%
First we need to calculate the theoretical continuous Fourier Transform (FT) of the discrete data, namely the discrete-time FT (DTFT)~\cite{byrne2015}. The 2D-DTFT of the 2D data $r_{\ell,k}$ is:
\begin{align}
	G(f_1,f_2)=\sum_{k}\sum_{\ell} r_{\ell,k} e^{-j2\pi \ell \frac{f_1}{f_{s_1}}}e^{-j2\pi k \frac{f_2}{f_{s_2}}}
	\label{eqn:2d-dtft}	
\end{align}
where $f_1,f_2$ are the frequencies and $f_{s_1},f_{s_2}$ are the corresponding sampling frequencies~\cite{oppenheim2010}. In our application the data $r_{\ell,k}$ are sampled spatially. Consequently, the frequencies and sampling rates have to correspond to the data sampled in the spatial domain. Our objective is to estimate the signal power from the estimated matrix $\mathbf{\hat{R}}$. Hence, we need to apply the DTFT by taking into account the signal model in~\eqref{eqn:ura-correlation-interferometer}.

\begin{figure}[t]
	\centering	
	\subfigure[signal of interest (SOI) at $\theta$=\ang{50}, $\phi$=\ang{10}.]{\includegraphics[width=0.99\linewidth]{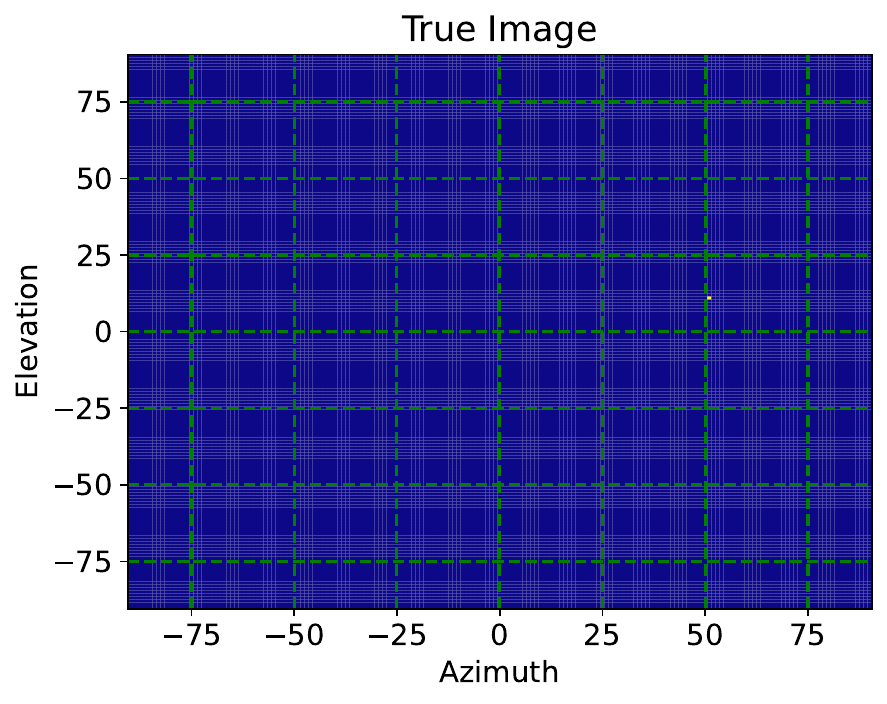}\label{fig:true-image-2d}} \hspace{-0.2cm}	
	\subfigure[2D-IDFT of the covariance matrix. ]{\includegraphics[width=0.99\linewidth]{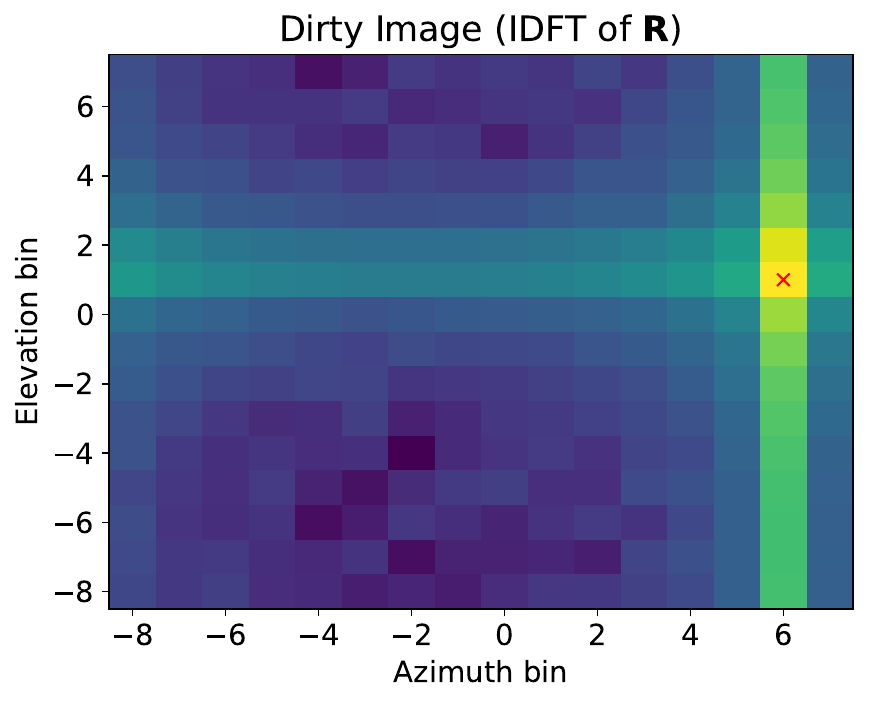}\label{fig:dft-2d-dirty}}
	\caption{(a) Real source location, (b) input image to the DNN which also includes the location of the look angle (the red x).}
\end{figure}

By simple inspection of \eqref{eqn:ura-correlation-interferometer} we notice that if we want the result of the DTFT of the data $r_{\ell,k}$ in \eqref{eqn:2d-dtft}	to correspond to the signal power $\sigma_x^2(\theta,\phi)$, we must replace the frequencies $f_1,f_2$ in \eqref{eqn:2d-dtft} with the spatial frequencies $\cos(\phi)\sin(\theta)/\lambda$ and $\sin(\phi)/\lambda$ respectively, and the sampling rates $f_{s_1},f_{s_2}$ with the spatial sampling rates $1/d_y$ and $1/d_z$ respectively. Taking then the 2D-DTFT over all the pairs of antennas we have that the estimated power of the signal over the continuous variables that indicate spatial direction $\phi,\theta$ is:
\begin{align}
	\hat{\sigma}^2_x(\theta,\phi)&=\sum_{k,\ell}  \frac{\hat{r}_{\ell,k}}{NM} e^{-j2\pi \ell \frac{\cos(\phi)\sin(\theta)/\lambda}{1/d_y}}e^{-j2\pi k \frac{\sin(\phi)/\lambda}{1/d_z}} \nonumber\\
	& \ell \in [0,M],~k \in [0,N].
	\label{eqn:ura-dft-correlation-interferometer}
\end{align}
%
\textbf{Remark 1:} This expression highlights the fact that with FT we are essentially doing beamforming of the correlations $r_{\ell,k}$ by setting as the BF coefficients the DTFT weights: In~\eqref{eqn:ura-dft-correlation-interferometer} we combine the correlation data coherently by correcting their phase offset due to the instrument geometry.

\textbf{2D-DFT:} Now we want to discritize the 2D-DTFT in \eqref{eqn:ura-dft-correlation-interferometer} so that we can calculate it in practice. We must derive the 2D-DFT with a number of $uFFT$, $vFFT$ points that are selected depending on the desired resolution. We index these DFT points with $u \in [\frac{-uFFT}{2},\frac{uFFT}{2}-1],v \in [\frac{-vFFT}{2},\frac{vFFT}{2}-1]$. The 2D-DFT of a set of 2D discrete data $r_{\ell,k}$ is then:
\begin{align}
	I[u,v]=\sum^{N-1}_{k=0}\sum^{M-1}_{\ell=0} r_{\ell,k} e^{-j2\pi \ell \frac{u}{uFFT}}e^{-j2\pi k \frac{v}{vFFT}}
	\label{eqn:2d-dft}	
\end{align}
\textbf{Remark 2:} A representative image of the 2D-DFT  after applying \eqref{eqn:2d-dft} can be seen in Fig.~\ref{fig:dft-2d-dirty} where a single point source 
is present. However, we must not forget that this DFT processing produces the \textit{dirty image} with the DTFT~\cite{thompson2017}. The term dirty image is used when the result is affected by the available DFT bins, cannot capture precisely a frequency component present in the signal, resulting in sidelobes.


\textbf{Conversion to Angular Bearings:} One final step has to do with the interpretation of these images. We convert the result of 2D-DFT beamforming in \eqref{eqn:2d-dft} to angles $\theta,\phi$ instead of using the binning of the 2D-DFT transform as seen in Fig.~\ref{fig:dft-2d-dirty}. For the DTFT to DFT conversion, we can see that the mapping from the continuous frequencies in~\eqref{eqn:2d-dtft} to the discrete in~\eqref{eqn:2d-dft} is accomplished by setting $\frac{f_1}{f_{s_1}}=\frac{u}{uFFT},\frac{f_2}{f_{s_2}}=\frac{v}{vFFT}$. The normalized spatial frequency over the sampling rate is between $1/2 \leq f_1/f_{s_1} \leq 1/2$. We have already substituted in \eqref{eqn:ura-dft-correlation-interferometer} the spatial frequencies $f_1,f_2$ that  are $\cos(\phi)\sin(\theta)/\lambda$ and $\sin(\phi)/\lambda$ respectively. Hence, from \eqref{eqn:ura-dft-correlation-interferometer}, \eqref{eqn:2d-dft} we simply solve the following: 
\begin{align*}
&\frac{f_2}{f_{s_2}}=\frac{v}{vFFT} \Rightarrow \frac{\sin(\phi)/\lambda}{1/d_z}=\frac{v}{vFFT} \nonumber \\
& \frac{f_1}{f_{s_1}}=\frac{u}{uFFT} \Rightarrow \frac{\cos(\phi)\sin(\theta)/\lambda}{1/d_y}=\frac{u}{uFFT}
\end{align*}
By solving the last two, the final angle locations are obtained. For the \textbf{elevation bin} $v$ the angle is
\begin{align}
\phi= \arcsin ( \frac{v}{nFFT} \frac{\lambda}{d_z}),
\end{align}
and for the \textbf{azimuth bin} $u$ the angle depends on the elevation:
\begin{align}
\theta= \arcsin ( \frac{u}{uFFT} \frac{\lambda}{d_y} \frac{1}{\cos(\phi)})~\textrm{rad}.
\end{align}


\begin{figure}[t]
	\centering
	\begin{subfigure}
		\centering
		\includegraphics[width=0.32\linewidth]{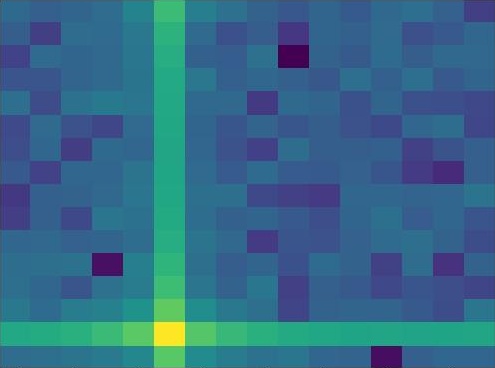}
	\end{subfigure}%
	\begin{subfigure}
		\centering
		\includegraphics[width=0.32\linewidth]{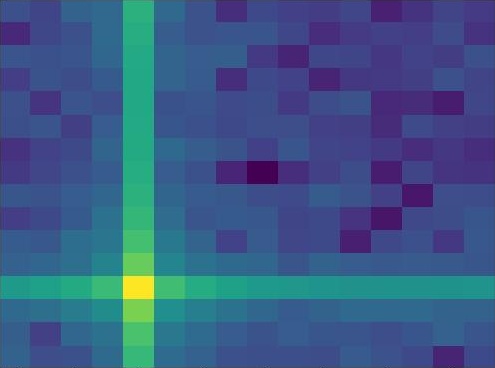}
	\end{subfigure}%
	\begin{subfigure}
		\centering
		\includegraphics[width=0.32\linewidth]{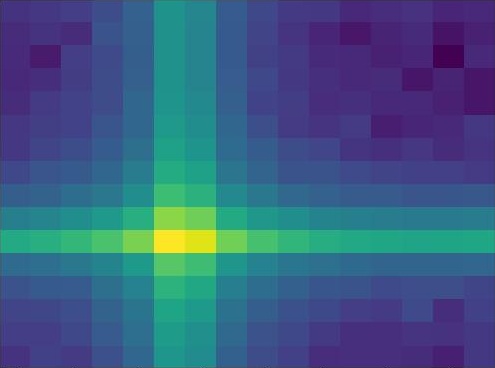}
	\end{subfigure}
	
	\caption{2D-DFT images with a moving jammer.} 
	\label{fig:dataset-no-jammer}
\end{figure}

\section{Dataset } The dataset was synthetically generated to simulate the URA responses for different SOI locations. It consists of the 2D-DFT images $I[u,v]$ (calculated in \eqref{eqn:2d-dft}). A typical image in the dataset is presented in Fig.~\ref{fig:dft-2d-dirty}. 
For producing the training dataset that contained only the SOI we varied its spatial location to cover the entire $M\times N$ 2D-DFT image. The SNR of the SOI varied from \SI{-9}{\decibel} to \SI{20}{\decibel} with a step size of \SI{1}{\decibel}. This approach generated $M\times N\times\mathrm{30}$ images. The training and validation datasets consisted of different perturbations of images in the time domain. 
The testing dataset consisted of clean images but also images that contain RFI and jammers. RFI and jamming sources can be \textit{transient} and appear in a single image, or \textit{moving} and \textit{static} that appear in more than one image. Fig.~\ref{fig:dataset-no-jammer} has an example of a moving jammer. The testing dataset consisted of RFI/jamming sources with an interference to noise ratio (INR) ranging from 0 to 30 dB.

\section{Sparsity-Inducing LSTM Autoencoder}
\begin{figure}[t]
	\centering
	\includegraphics[width=0.99\linewidth]{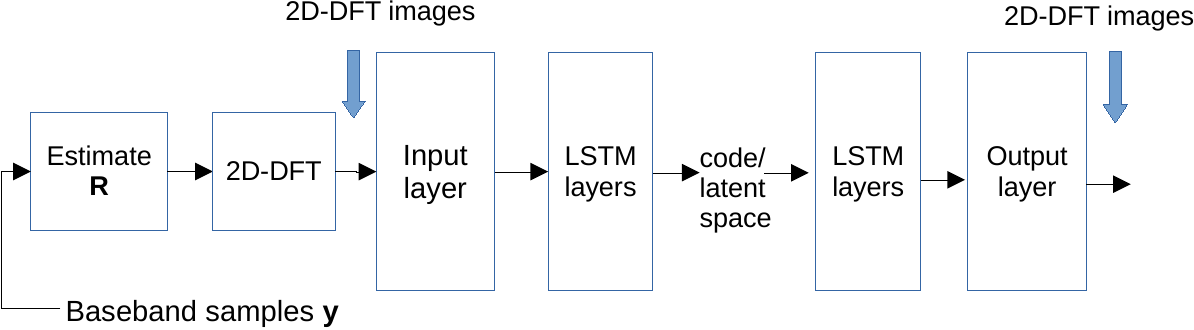}
	\caption{LSTM autoencoder with 2D-DFT image inputs.}
	\label{fig:lstm-autoencoder}
\end{figure}

The neural network (NN) model used in this work is an LSTM autoencoder. The autoencoder compresses the input data into a condensed representation, which it then reconstructs it to its original form (Fig.~\ref{fig:lstm-autoencoder}).
We will now write formally the operations of the auto-encoder for the 2D-DFT of matrix $\mathbf{R}$ that was calculated from our spatially-sampled data. The input vector denoted as $\mathbf{i}$, consists of the data given by \eqref{eqn:2d-dft}. Its reconstruction denoted as $\mathbf{\hat{i}}$. The output of the encoder is the latent representation $\mathbf{h}$ which is also referred to as the \textit{code}~\cite{goodfellow2016}. $\mathbf{W}_E,\mathbf{W}_D$ are the parameters of the encoder and decoder respectively that are trained, and $\mathcal{L}$ is the loss function that is the MSE plus the sparsity loss:
\begin{align}
\mathbf{h}&=f_\text{E}(\mathbf{i},\mathbf{W}_E),~~
\mathbf{\hat{i}}=f_\text{D}(\mathbf{h},\mathbf{W}_D),\nonumber\\\
\mathcal{L}&=\frac{1}{T}\sum_{t}  \Big ( \|\mathbf{i}^{(t)}-\mathbf{\hat{i}}^{(t)} \|^2_2+\alpha\|\mathbf{h}^{(t)}\|_1\Big ).
\end{align}
The number of training batches is $T$, $\| \cdot\|_2$ is the L2 norm, $\alpha$ is the hyperparameter that determines the weight of the sparsity-inducing L1 norm~\cite{goodfellow2016}. 
Contrary to image classification with a convolutional NN (CNN), our input observation consists of a sequence of $P$ successive images of size $M$$\times$$ N$. In this work, the input layer has as input $M$$\times$$N$=$128$ $\times$$128$ features, and $P$=10. Hence, we detect RFI/jamming based on events occurring in a sequence of $P$ images and not a single image.


The NN architecture implemented in this case was comprised of 3 LSTM layers at the encoder $f_\text{E}$, and 3 at the decoder $f_\text{D}$. The input/output dimensions of the three LSTM layers are 128x128/64x64, 64x64/32x32, 32x32/16x16, i.e., the number of nodes in the code layer (code size) is 16x16. The resulting LSTM autoencoder model has $\approx$ 23M parameters. The Adam optimizer was used as the solver during training with a batch size of 32 observations. 
We fine-tuned the learning rate to 0.02 and the sparsity weight to $\alpha=0.001$.

\section{Results}
\begin{figure}[t]
	\centering
\subfigure[]{	\includegraphics[width=0.85\linewidth]{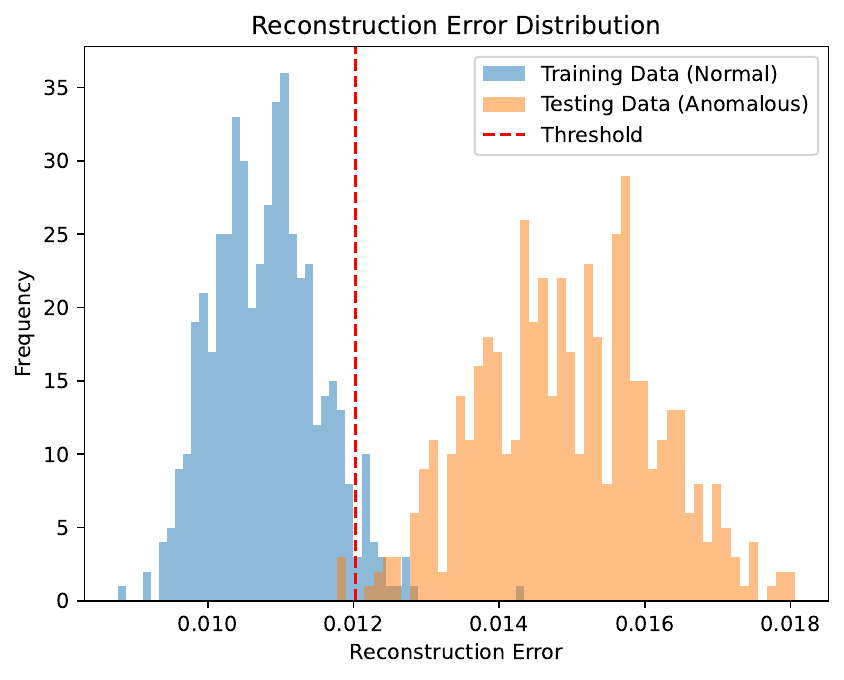}}	
\subfigure[]{	\includegraphics[width=0.85\linewidth]{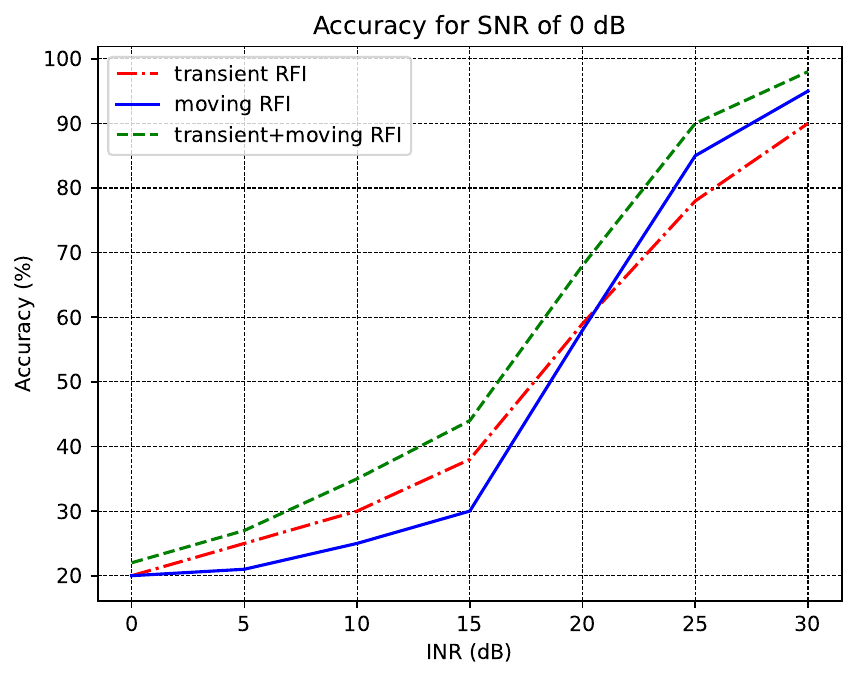}}
	
	\caption{(a) Distribution of the error after image reconstruction with the auto-encoder. (b) Accuracy vs. INR for different number and types of RFI sources in the testing dataset.}
	\label{fig:results1}
\end{figure}

After training, the model is used for anomaly detection~\cite{malhotra2015}. Detection of RFI or jamming is declared when, during testing, the reconstruction error of the input sequence of $P$ images is more than a threshold (more than the 95th percentile of the training reconstruction error).
In Fig.~\ref{fig:results1}(a) we show the distribution of the reconstruction error of the autoencoder for an INR of 20dB and a very low SOI power with an SNR of 0 dB. The anomalous data are easily distinguishable even for this low SOI power. In Fig.~\ref{fig:results1}(b) we present the detection accuracy for different values of the RFI INR. We also explore different types of RFI/jamming sources. In the low INR regime, the RFI source(s) are very weak and so their presence cannot be easily detected. As the INR is increased, we can detect them better despite the low power SOI. Notice that the most challenging case is the transient RFI/jammer. Note also that as moving jammers are added, we have a performance increase due to the image being more distinguishable even for low INR. Due to lack of space, we do not present a SNR higher than 0 dB since it is more favorable for our system as expected.

%


\section{Conclusions}
\label{section:conclusions}
In this paper, we presented a new method to detect RFI or jamming with DNNs but using as input features the spatially-sampled antenna array data that are converted to images. This is contrary to spectrograms and time-domain baseband samples that are used as input in existing RFI and jamming classification schemes. Our performance evaluation suggests that this data modality can lead to high classification accuracy of RFI or jamming when we have antenna array systems.


\bibliographystyle{IEEEtran} 
\bibliography{../../../MyLibrary}

\end{document}